\documentclass[sigconf]{acmart}
\AtBeginDocument{%
  \providecommand\BibTeX{{%
    \normalfont B\kern-0.5em{\scshape i\kern-0.25em b}\kern-0.8em\TeX}}}

\setcopyright{acmlicensed}
\copyrightyear{2018}
\acmYear{2018}
\acmDOI{XXXXXXX.XXXXXXX}

\acmConference[Conference acronym 'XX]{Make sure to enter the correct
  conference title from your rights confirmation emai}{June 03--05,
  2018}{Woodstock, NY}
\acmISBN{978-1-4503-XXXX-X/18/06}

\usepackage{enumitem}
\begin{document}

\title{Deep Group Interest Network on Full Lifelong User Behaviors for CTR Prediction}

\author{Qi Liu}
\affiliation{%
  \institution{University of Science and Technology of China}
  \city{Hefei}
  \country{China}
}
\email{qiliu67@mail.ustc.edu.cn}

\author{Xuyang Hou}
\affiliation{%
  \institution{Meituan}
  \city{Beijing}
  \country{China}
}
\email{houxuyang@meituan.com}

\author{Haoran Jin}
\affiliation{%
  \institution{University of Science and Technology of China}
  \city{Hefei}
  \country{China}
}
\email{haoranjin@mail.ustc.edu.cn}

\author{Xiaolong Chen}
\affiliation{%
  \institution{University of Science and Technology of China}
  \city{Hefei}
  \country{China}
}
\email{chenxiaolong@mail.ustc.edu.cn}

\author{Jin Chen}
\affiliation{%
  \institution{University of Electronic Science and Technology of China}
  \city{Chengdu}
  \country{China}
}
\email{chenjin@std.uestc.edu.cn}

\author{Defu Lian}
\affiliation{%
  \institution{University of Science and Technology of China}
  \city{Hefei}
  \country{China}
}
\email{liandefu@ustc.edu.cn}

\author{Zhe Wang}
\affiliation{%
  \institution{Meituan}
  \city{Beijing}
  \country{China}
}
\email{wangzhe65@meituan.com}

\author{Jia Cheng}
\affiliation{%
  \institution{Meituan}
  \city{Beijing}
  \country{China}
}
\email{jia.cheng.sh@meituan.com}

\author{Jun Lei}
\affiliation{%
  \institution{Meituan}
  \city{Beijing}
  \country{China}
}
\email{leijun@meituan.com}

\renewcommand{\shortauthors}{Trovato and Tobin, et al.}

\begin{abstract}

Modeling user interest based on lifelong behavior sequences is key for improving Click-Through Rate (CTR) predictions. Current approaches typically use a two-step process to balance efficiency with effectiveness. Initially, they use an effective algorithm to identify relevant historical behaviors to candidates in the first phase, and then they focus on a shorter subsequence to ascertain user interest via target attention. However, this two-step approach, despite its effectiveness, unavoidably results in some loss of information. Moreover, limiting interest modeling to just click behaviors introduces bias, as other historical behaviors like purchases also affect the likelihood of a click. These issues prevent the CTR prediction model from achieving its full potential. In our study, we introduce the \textbf{D}eep \textbf{G}roup \textbf{I}nterest \textbf{N}etwork (\textbf{DGIN}), a comprehensive method that processes the entire spectrum of a user's lifelong behavior, including clicks, collections and purchases, in an end-to-end manner. We start by organizing the complete lifelong behavior sequences into groups based on a specific interest key, significantly reducing the sequence length from tens of thousands to hundreds. To overcome the potential information loss from this grouping, we make the following two designs. Firstly, we analyze behaviors within each group using both simple statistics and self-attention to capture group traits and then pinpoint user interests by applying target attention to these groups. Secondly, we refine the user's decision-making interest by employing the attention mechanism to identify the user's candidate-specific interests, based on behavior subsequences that share the same interest key. Our extensive experiments on both industrial and public datasets confirm the effectiveness and efficiency of DGIN. The A/B test in our LBS advertising system shows that DGIN improves CTR by 4.5\% and Revenue per Mile by 2.0\%.

\end{abstract}






\maketitle

\section{Introduction}
\begin{figure}
    \centering
    \includegraphics[width=\linewidth]{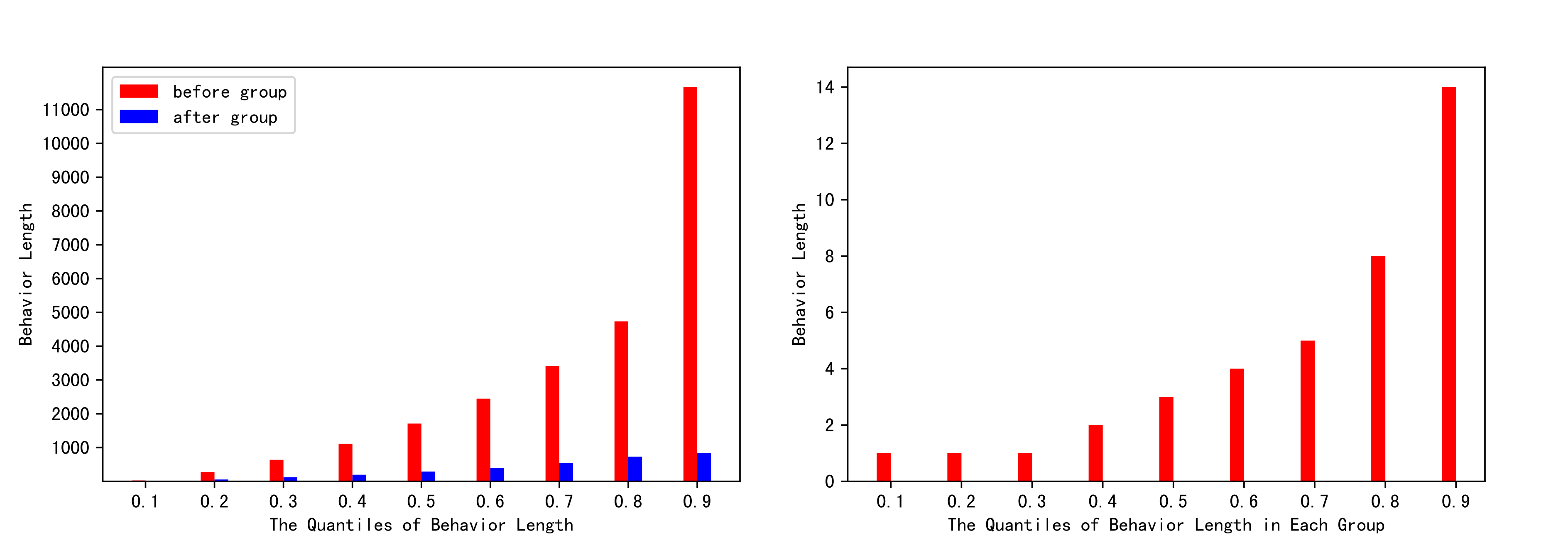}
     \vspace{-1em}
    \caption{The distribution of behaviors' length affected by grouping on \emph{item\_id}. We rank samples on behaviors' length in ascending order. The left half shows the corresponding behavior's length at different quantiles. The right half displays the distribution of the behavior quantity of each group.}
    \label{fig:motivation_num_hist}
    \vspace{-1em}
\end{figure}
%In modern recommendation systems (RS), Click-Through Rate (CTR) prediction serves a vital role in estimating the probability of a user clicking on candidate items. The estimated probability is a decisive factor in the final ranking formula, which determines the exposed items to the user. In recent years, numerous approaches have been proposed to improve the accuracy of CTR prediction. Mining interest patterns from the user's historical behaviors is a main research topic. Some works~\cite{pi2019practice,pi2020search} observed that lifelong user behavior sequences were more beneficial than the most recent ones since they describe user comprehensive preferences more accurately. However, the high demands of online systems for low latency pose significant challenges to the effective processing of lifelong behavior sequences. It's urgent to unlock the potential of lifelong behavior sequences.
%the strict online latency limitation poses challenges in processing lifelong behavior sequences effectively.

In contemporary recommendation systems, predicting the Click-Through Rate (CTR) is crucial for determining the likelihood of a user engaging with suggested items. This prediction significantly influences the algorithm that ranks and presents items to users. Over recent years, there has been a surge in methods aimed at improving CTR prediction accuracy, with a significant focus on extracting patterns of interest from users' past activities. Studies have shown that analyzing sequences of user behavior over their entire history is more advantageous than just considering recent actions, as it offers a more complete picture of their preferences. However, the need for low latency in online platforms presents a considerable obstacle to efficiently processing these extensive behavior sequences. Therefore, there's a pressing need to explore the full potential of these lifelong behavior sequences for improved recommendation accuracy.

%Existing works on lifelong behavior sequence modeling typically follow a two-stage approach. Firstly, a fast and lightweight retrieval module identifies hundreds of historical behaviors related to the candidate item from the tens of thousands of lifelong behaviors. Secondly, target attention is applied to deduce the user's interest from the retrieved subsequence. The focus of research has primarily been on the first stage. SIM~\cite{pi2020search} retrieves behaviors of the same category\_id or cluster as the candidate item. UBR4CTR~\cite{qin2020user} applies the BM25~\cite{robertson2009probabilistic} as a relevance score and uses the inverted index to obtain relevant behaviors. ETA~\cite{chen2022efficient} utilizes locality-sensitive hashing (LSH)~\cite{datar2004locality} and Hamming distance to perform user behavior retrieval based on low-cost bit-wise operations. SDIM~\cite{cao2022sampling} selects behaviors that share the same hash signature as the candidate item by multi-round hash collision. TWIN~\cite{chang2023twin} solves the relevance score inconsistency problem between two stages by using shared efficient target attention. The two-stage paradigm represents a trade-off between effectiveness and efficiency.

Existing approaches to modeling lifelong behavior sequences typically adopt a two-stage methodology. Initially, a rapid and lightweight retrieval module sifts through hundreds of historical behaviors pertaining to the candidate item from a vast pool of lifelong behaviors. Subsequently, target attention is employed to infer the user's interest based on the retrieved subsequence. The primary focus of research has predominantly revolved around the initial stage. For instance, SIM~\cite{pi2020search} retrieves behaviors belonging to the same category or cluster as the candidate item. UBR4CTR~\cite{qin2020user} employs BM25~\cite{robertson2009probabilistic} for relevance scoring and utilizes an inverted index for retrieving relevant behaviors. ETA~\cite{chen2022efficient} employs locality-sensitive hashing (LSH)~\cite{datar2004locality} for efficient user behavior retrieval. SDIM~\cite{cao2022sampling} selects behaviors sharing the same hash signature as the candidate item through multi-round hash collision. TWIN~\cite{chang2023twin} addresses the inconsistency in relevance scoring between the two stages by employing shared efficient target attention. This two-stage approach embodies a compromise between effectiveness and efficiency.

%Although effective, those algorithms still suffer two non-negligible limitations: biased and incomplete interest estimation caused by retrieval-based modeling and the sole use of historical click behaviors. The biased estimation of interest results from the first stage retrieval, which retains only the top-relevant behaviors and discards at least $95\%$ of the historical behaviors. The incomplete estimation is caused by the use of the historical click behavior to form the lifelong behavior sequence, out of the concern about the consistence with the CTR task. However, there are many different types of interactions between users and items. For example, in an online location-based services (LBS) platform, the behaviors of interaction range from click, browse-dishes, and view-comments to add-to-cart and purchase. The lifelong click behavior sequence alone cannot comprehensively depict the user's interest since the intensity of interest reflected by different types of behaviors varies~\cite{grbovic2018real}. Thus, the full lifelong user behaviors provide an opportunity to capture fine-grained interest patterns. 

While effective, these algorithms still face two significant limitations: biased and incomplete interest estimation stemming from retrieval-based modeling and reliance solely on historical click behaviors, respectively. Biased interest estimation arises from the retrieval process in the first stage, which retains only the most relevant behaviors while discarding at least $95\%$ of historical behaviors. Incomplete interest estimation results from the only use of historical click behaviors to construct the lifelong behavior sequence, driven by concerns about consistency with the Click-Through Rate (CTR) task. However, there exists a wide array of user-item interactions. For example, in an online location-based services (LBS) platform, the behaviors of interaction range from click, browse-dishes, and view-comments to add-to-cart and purchase. Relying solely on lifelong click behavior sequences cannot adequately depict a user's interests since the significance of interest varies across different behavior types~\cite{grbovic2018real}. Hence, utilizing the entirety of lifelong user behaviors presents an opportunity to capture fine-grained interest patterns comprehensively.

%To address the bias and incomplete interest estimation, we propose \textbf{D}eep \textbf{G}roup \textbf{I}nterest \textbf{N}etwork (\textbf{DGIN}) for full lifelong user behavior modeling. DGIN extracts the user's interest from the full lifelong behavior sequence in an efficient end-to-end manner. The relationship between interest and behavior is that interest determines behavior, and behavior reflects interest. While there may be unlimited behaviors, there are only limited interests. Therefore, we first organize full lifelong behaviors into interest groups based on an interest key. This key acts as the interest center, which can be an existing concept (e.g. category\_id, item\_id) or learned from the data. The grouping operation transforms the lifelong behavior sequence into interest groups, such that the magnitude of sequence length is significantly reduced. As shown in Figure~\ref{fig:motivation_num_hist}, the grouping operation reduces the magnitude of behavior from $O(10^4)$ to $O(10^2)$. In our approach, we use the item\_id as the key for grouping, according to the observation that users have repeated consumption habits in our online LBS platform. As shown in the experiment part, we find that grouping by category\_id also leads to performance gains. Considering category information exists almost in all recommendation scenarios, the proposed approach is universal. 

To tackle the issues of biased and incomplete interest estimation, we introduce the Deep Group Interest Network (DGIN) for comprehensive modeling of lifelong user behaviors. DGIN efficiently extracts user interests from the entirety of lifelong behavior sequences in an end-to-end manner. It is well-known that interests play a significant role in shaping behavior, and behavior often provides clues about underlying interests. Therefore, the rate of growth in the number of behaviors over time exceeds the ratio of the number of interests by orders of magnitude. Consequently, we initially organize the full lifelong behaviors into interest groups based on a designated interest key. This key serves as the focal point of interest and can either be a predefined concept (e.g., category\_id, item\_id) or learned from the data. The grouping operator transforms the lifelong behavior sequence into interest groups, significantly reducing the behavior length. Illustrated in Figure~\ref{fig:motivation_num_hist}, this grouping operation diminishes the magnitude of behavior length from $O(10^4)$ to $O(10^2)$. In our methodology, we utilize the item\_id as the grouping key, considering the recurrent consumption habits observed in our online Location-Based Services (LBS) platform. Additionally, our experiments demonstrate performance improvements when grouping behavior sequences by category\_id. Given that category information is prevalent across various recommendation scenarios, our proposed approach exhibits universality.

To mitigate the loss of information resulting from behavior grouping, we make the following two designs. First, we conduct an analysis of the behaviors within each group by utilizing both simple statistics and self-attention mechanisms to identify the distinctive traits of each group, and then we apply target attention to these groups to accurately identify user interests. Specifically, we calculate statistics, such as the frequency of different behaviors within a group, to gauge the user's level of interest. Simultaneously, we employ self-attention on each group of behaviors to capture its unique characteristics, since the details of the behaviors (like the type of behavior and the time it occurs) reveal how a user's interests evolve over time. For instance, a user might only buy coffee during weekdays, not weekends. This approach aims to retain as much unique behavior information within each group as possible. The subsequent use of target attention is to precisely determine the user's interests from these interest groups. This comprehensive approach reduces the information loss. Second, the complete history of a user's behavior reveals patterns of how their interests evolve towards certain items, which can help predict how likely they are to click on the item. To do this, we use self-attention to deduce interests specific to a candidate item from subsequences of behaviors that align with the candidate's interest key. We then refine this with target attention, allowing us to understand the user's decision-making process towards the candidate, particularly in Location-Based Services (LBS) platforms where repeated interactions are common. Overall, we make the following contributions:
\begin{itemize}[leftmargin=*]
    \item We are the first to achieve efficient end-to-end interest extraction by taking all behaviors into calculation in lifelong/long behavior sequence modeling. We reveal the necessity of introducing multiple types of behaviors into the lifelong sequence modeling. 
    \item We propose a Deep Group Interest Network for capturing the user's long-term interest, where we organize lifelong behavior sequences into interest groups, remarkably reducing computation overload. We also sample a subsequence to capture the user's decision pattern towards the candidate.
    \item To evaluate the effectiveness of DGIN, we conduct offline experiments on both industrial and public datasets. The results demonstrate a remarkable improvement achieved by DGIN. The A/B test in our LBS advertising system shows DGIN improves CTR by 4.5\% and Revenue per Mile by 2.0\%.
\end{itemize}

\section{Related Work}
\subsection{Click-Through Rate Prediction}
CTR prediction has been a long-standing research hotspot in RS. Early CTR methods~\cite{wright1995logistic,rendle2010factorization,juan2016field} mainly focus on the low-order feature interactions. Recently, methods based on deep learning have achieved amazing progress in CTR prediction. Wide\&Deep~\cite{cheng2016wide} utilizes the linear model to memorization of feature interaction and takes the deep neural network to achieve generalization. DeepFM~\cite{guo2017deepfm} replaces the linear model with FM to emphasize the second-order feature interactions. DCN~\cite{wang2017deep} applies a cross-vector network to learn informative feature interactions automatically. 

User behavior sequence modeling~\cite{zhou2018deep,feng2019deep,zhou2019deep,chen2019behavior,gu2020deep,xu2020deep,zhao2022non} also attracts lots of attention and develops fast. It focuses on extracting the user's interest from historical behaviors to improve CTR prediction accuracy. Limited by the online latency, most existing methods design algorithms on the truncated short user behavior sequence, which only contains the user's instant interest. DIN~\cite{zhou2018deep} first performs attention between candidate item and behavior sequence to extract interest by emphasizing candidate-relevant behaviors and suppressing candidate-irrelevant ones. DIEN~\cite{zhou2019deep} further takes a two-layer GRU~\cite{chung2014empirical} to model the temporal shifting and mine interest in the interest level. DSIN~\cite{feng2019deep} divides behavior sequence into sessions and uses self-attention together with Bi-LSTM~\cite{graves2005framewise} to obtain session representation. Then it extracts interest from sessions' representation through target attention. However, when applying DSIN to lifelong behavior sequence modeling, the setting of the session's time interval is troublesome. When the time interval is small, there are lots of sessions leading to inefficiency. When the time interval is large, there will many heterogeneous behaviors that hold different item\_ids and different category\_ids within each session and the session aggregation will cause severe information loss. NINN~\cite{zhao2022non} just partition the behavior sequence into different categories and capture the interactions among them. However, it does not supplement any statistical or dynamic information, which will give rise to performance degradation. Meanwhile, some works introduce multiple types of behavior sequences~\cite{zhou2018atrank,zhou2018micro,ouyang2019deep,guo2019buying,gu2020deep,xie2021deep} into CTR modeling to obtain comprehensive fine-grained interest. ATRANK~\cite{zhou2018atrank} utilizes the Tansformer~\cite{vaswani2017attention} encoder-decoder network to acquire interest from the mixed behavior sequence which consists of various types of behaviors in chronological order. While DMT~\cite{gu2020deep} uses different behavior modeling networks to process different types of behavior sequences. However, the mere use of the short behavior sequence can not obtain important long-term interest patterns and restrict the performance improvement of CTR prediction. 

\subsection{Long User Behavior Sequence Modeling}
Due to the effectiveness of user behavior sequence modeling, long behavior sequences modeling~\cite{pi2019practice,pi2020search,chen2022efficient,cao2022sampling,zhang2022clustering,qin2023learning,yang2023empowering} have been explored. MIMN~\cite{pi2019practice} applied the memory network and GRU to induce interest stored at the user interest center. However, MIMN barely processes sequences longer than $10^3$, and the induction process without the candidate item causes lots of information loss. After MIMN, the two-stage solution becomes the mainstream. The first stage retrieves hundreds of candidate-relevant behaviors from the long behavior sequence in an efficient way and then performs target attention with the retrieved behaviors to extract interest. SIM Hard~\cite{pi2020search} takes behaviors that have the same category as the candidate item. SIM Soft~\cite{pi2020search} selects top-ranked behaviors according to the inner product between the candidate item and historical behaviors with pre-trained item embeddings. UBR4CTR~\cite{qin2020user} chooses the BM25 as the retrieve metric. Recently, ETA~\cite{chen2022efficient} and SDIM~\cite{cao2022sampling} have tried to do retrieving and extracting in an end-to-end manner. ETA~\cite{chen2022efficient} uses LSH to perform item embedding binarization and takes the Hamming distance as the metric to filter behaviors. SDIM~\cite{cao2022sampling} selects behaviors that hold the same hash signature as the candidate item through multi-round hash collision and the embedding of selected behaviors will be aggregated directly. TWIN~\cite{chang2023twin} solves the relevance inconsistency problem by sharing an efficient target attention network between two stages. Although end-to-end, the interest is still extracted from the retrieved subsequence and can't escape the problem of biased interest. All those methods only retrieve candidate-relevant subsequence from the click behaviors, which leads to biased and incomplete interest modeling. The above problems motivate us to explore an efficient end-to-end method of taking all information for full lifelong user behavior sequence modeling. Unlike existing ETA, SDIM, and TWIN, which adhere to a two-stage end-to-end framework where only the candidate-relevant behaviors contribute to the gradient, DGIN takes a different path. DGIN utilizes a grouping strategy, enabling all behaviors in the lifelong behavior sequence to actively participate in extracting interest. This design ensures that each gradient update is associated with all behaviors during the backward process, facilitating end-to-end training with full information. 

\section{Methods}

\subsection{Preliminaries}
\begin{figure*}
    \centering    \includegraphics[width=\textwidth]{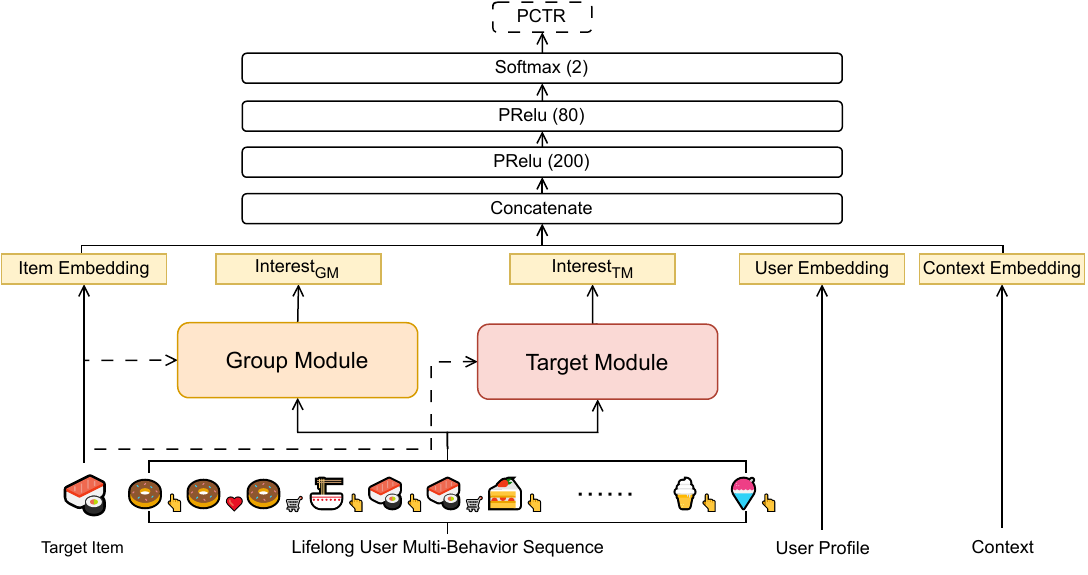}
    % \vspace{-1em}
    \caption{The overall framework of Deep Group Interest Network (DGIN). DGIN consists of Group Module (GM) and Target Module (TM), which capture long-term and decision interest from lifelong user multi-behavior sequence.}
    \label{fig:dgin}
    % \vspace{-1em}
\end{figure*}
CTR prediction aims at estimating the probability of the user clicking a candidate item under a specific context in the ranking stage. The instance can be represented by $(\mathbf{x}, y)$, where $\mathbf{x}=[\mathbf{x}^u, \mathbf{x}^s, \mathbf{x}^i, \mathbf{x}^c]$, $y\in\{0,1\}$ indicates click or not. $\mathbf{x}^{u}, \mathbf{x}^{s}, \mathbf{x}^i$, and $\mathbf{x}^c$ represent the features' set of user, user behavior sequence, candidate item, and context respectively. Given training dataset $D=\{(\mathbf{x}_1, y_1),...,(\mathbf{x}_N, y_N)\}$, we need to learn a model $f$ to predict the CTR, which can be formulated as the following Eq.~(\ref{eq:ctr_formulation}):
\begin{equation}
    \label{eq:ctr_formulation}
    p_i=f(\mathbf{x}).
\end{equation}
where $p_i$ is the estimated probability and $f$ is the CTR model.
CTR model is usually trained as a binary classification problem by minimizing the negative log-likelihood loss on the training dataset:
\begin{equation}
    \label{eq:ctr_loss}
    l(D)=-\frac{1}{N} \sum_{i=1}^{N}y_i \log (p_i)+(1-y_i)\log (1-p_i).
\end{equation}
where $N$ is the size of the training dataset. For conciseness, we omit the subscript $i$ in the following description when no confusion. 

As shown in Figure~\ref{fig:dgin}, DGIN is composed of the Embedding layer, the Group Module (GM), the Target Module (TM), and Multi-Layer Perception (MLP). The original input $\mathbf{x}$ of the CTR model is sparse one-hot vectors. It will go through the embedding layer to get the low-dimensional dense vectors as representation. The GM takes the already grouped lifelong behavior sequence as input. It first applies the self-attention mechanism to aggregate the unique characteristics of each behavior within the group and then performs target attention to extract user interest in an end-to-end manner. Meanwhile, the TM takes the subsequence whose behaviors hold the same interest key as the candidate item and then obtains the psychological decision interest towards the candidate through another attention network. The extracted long-term interest and psychological decision interest will serve as the input of the MLP.

\subsection{Embedding Layer} 
Each field has its own embedding matrix $\mathbf{E}=[\mathbf{e}_1;\mathbf{e}_2;\cdots;\mathbf{e}_K] \in \mathbb{R}^{K \times d}$, where $K$ represents the cardinality of the field and $d$ donates the embedding dimensional. The $\mathbf{e}_i$ severs as the embedding of the feature assigned index $i$ in the field. We make all fields share the same embedding dimensional. 

Since DGIN focuses on full lifelong user behavior sequence modeling, we provide processing detail about $\mathbf{x}^s$. The lifelong user behavior sequence consists of the user's various interactions (e.g. click, add-to-cart, browse-dishes, etc) with items in chronological order after registration. There $\mathbf{x}^s$ can be represented as $\mathbf{x}^s=[\mathbf{x}^s_1, \mathbf{x}^s_2, ..., \mathbf{x}^s_L]$, where $L$ is the length of the lifelong behavior sequence. To fully describe each behavior, each $\mathbf{x}^s_i$ has lots of attributes, such as \emph{item\_id, category\_id, price, timestamp, location, behavior\_type} etc. The embedding layer transform each attribute into corresponding embeddings $\{\mathbf{e}^s_{i,item\_id},..., \mathbf{e}^s_{i,timestamp}, ...,\mathbf{e}^s_{i,behavior\_type}\}$. We concatenate them together to form the behavior representation $\mathbf{e}^s_i=[\mathbf{e}^s_{i,item\_id},..., \mathbf{e}^s_{i,timestamp}, ..., \mathbf{e}^s_{i,behavior\_type}]$.

\subsection{Group Module}

\begin{figure}
    \centering
    \includegraphics[width=1.0\linewidth]{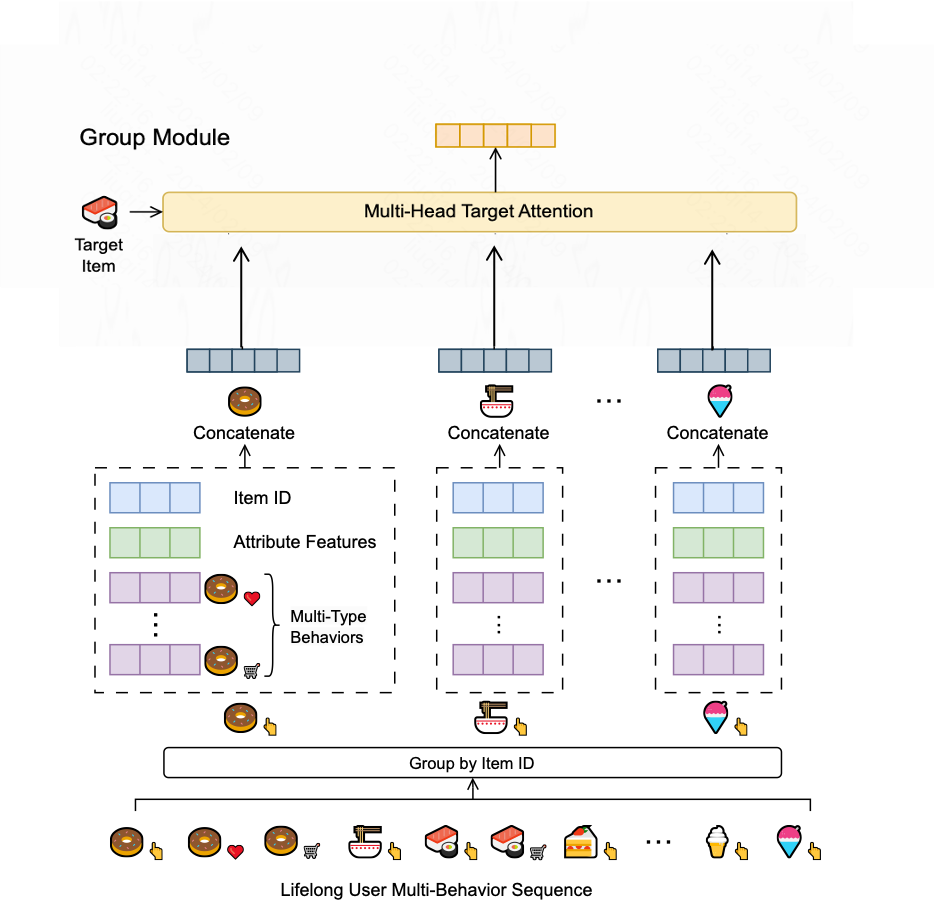}
    %\vspace{-1em}
    \caption{The detail of GM's architecture. We take item\_id as the interest key for illustration.}
    \label{fig:gm}
    %\vspace{-1em}
\end{figure}

Figure~\ref{fig:gm} shows the details of the Group Module. As the goal is to extract the user's interest towards the candidate item from the lifelong behavior sequence, it is reasonable to cluster the behaviors in a coarse interest level and then mine interest from the clustered interests, which will be more computationally efficient. In the offline data processing stage, GM first groups the lifelong behavior sequence into interest groups based on interest key. Within each group, the behavior subsets are still stored in chronological order. However, grouping behaviors will damage the information integrality such as the occurrence times of different types of interaction, the spatio-temporal relationship among behaviors, etc. To make up for the lost information, we designed two types of features as the attributes of the interest groups, including the statistical and aggregated attributes. The statistical attributes are statistics of behaviors within the group from different perspectives and the aggregated attributes are obtained by applying self-attention to the unique attributes of original behaviors, such as timestamp. The common attributes (e.g. item\_id, category\_id) shared by behaviors within the group together with the supplementary attributes form the attributes of the interest groups. Due to the limited interest groups, we employ Multi-Head Target Attention (MHTA) to extract unbiased and comprehensive interest from the interest groups.

\subsubsection{Statistical Attributes}
Inspired by \textbf{Quantity Breeds Quality}, we calculate the various statistics about quantity within the group. More behaviors towards the item mean more preference. Within the group, we count the total number of behaviors, total behavior types, and individual numbers of different types of behaviors. From the duration perspective, we calculate the average dwell time of all behaviors. And we will also obtain the average consumption amount of all purchase behaviors. The quantity, time, and money are three straightforward factors which reflect the user's interest. All the statistical attributes are processed during the offline data processing stage, and there is no extra inference cost. We represent the statistical attributes as Eq.~(\ref{eq:stat_attr}).
\begin{equation}
    \label{eq:stat_attr}
    \mathbf{attr}_s=[\mathbf{attr}_{counts},\mathbf{attr}_{types},...,\mathbf{attr}_{avg\_price}]
\end{equation}

\subsubsection{Aggregated Attribute}
Statistical attributes only reflect interest intensity, but can not tell the interest evolution. Within the group, the spatio-temporal attributes like \emph{timestamp, location} of behavior can represent the user's interest evolution process on the coarse interest. Meanwhile, we can also observe the user's attitude about the item from the heterogeneity interactions. For example, the recent click behavior may strongly impact the current click, while clicks long ago have little influence. However, purchases long ago may have a greater impact on current clicks. Thus, we use the attributes' sequence: \emph{timestamp, location, and behavior\_type} to supplement the information on the interest dynamics. 

Specifically, suppose there are maximum of $B$ behaviors in each group as $\mathbf{b}=<\mathbf{b}_{1}, \mathbf{b}_{2}, ..., \mathbf{b}_{B}>$. We concatenate the embeddings of \emph{timestamp, location, and behavior\_type} as the behavior's representation $\mathbf{e}_{b_i}=[\mathbf{e}_{b_i,timestamp},\mathbf{e}_{b_i,location},\mathbf{e}_{b_i,behavior\_type}]$. The intra-group behavior sequence can be expressed as $\mathbf{e}_b=[\mathbf{e}_{b_1}, \mathbf{e}_{b_2}, ..., \mathbf{e}_{b_B}]$. Due to the ability to model behavior pairs from multiple perspectives, we apply Multi-Head Self Attention MHSA to capture the interest evolution. The MHSA can be expressed as follows:
\begin{equation}
    MHSA(\mathbf{e}_b)=concat(head_1,...,head_h)W^O,
\end{equation}
\begin{equation}
    head_i=Softmax(\frac{\mathbf{e}_bW^Q_i({\mathbf{e}_bW^K_i})^T}{\sqrt{d'}})\mathbf{e}_bW^V_i,
\end{equation}
where h is the number of heads, $W^Q_i, W^K_i, W^V_i \in R^{3d \times d'}$, $W^O \in R^{3d \times 3d}$. $3d$ and $d'$ are the dimension of the input and weight vectors while $d'=\frac{3d}{h}$. Then, mean pooling is taken to process the $MHSA(\mathbf{e_b})$ and acquire the aggregated attribute as Eq.~(\ref{eq:agg_attr}).
\begin{equation}
    \label{eq:agg_attr}
    \mathbf{attr}_a=mean\_pool(MHSA(\mathbf{e}_b)).
\end{equation}
where $\mathbf{attr_a} \in R^{3d}$ is the aggregated attribute. 

\subsubsection{Attention On Interest Set}
The grouping operation changes the full lifelong behavior sequence into limited coarse interest groups. Three types of attributes are utilized to describe the characteristics of each interest group. The first is the identity attributes $\mathbf{attr_i}$ including interest\_key\_id, category\_id, etc. Secondly, statistical attributes are calculated to represent the interest intensity. Finally, aggregated attribute reflects the interest dynamic. After obtaining all attributes, we exploit the MHTA to perform interest activation to get the user's interest towards each candidate item. MHTA holds the network structure that takes the candidate item as query and the interest sets as key and value. In the ideal attention mechanism, the identity attributes (like category\_id) of different behaviors holding the same interest key will participate in the calculation of attention score many times, which will give rise to redundant computation. GM avoids lots of redundant computation and achieves computation efficiency.

Specifically, suppose there are maximum of $G$ interest groups in each full lifelong behavior sequence as $\mathbf{g}=<\mathbf{g}_{1}, \mathbf{g}_{2}, ..., \mathbf{g}_{G}>$. The representation of each interest group is $\mathbf{e}_{\mathbf{g}_i}=[\mathbf{e}_{\mathbf{attr}_i},\mathbf{e}_{\mathbf{attr}_s},\mathbf{e}_{\mathbf{attr}_a}]$. We can get the interest groups' representation $\mathbf{e}_\mathbf{g}=[\mathbf{e}_{\mathbf{g}_1}, \mathbf{e}_{\mathbf{g}_2}, ..., \mathbf{e}_{\mathbf{g}_G}]$. The fine-grained interest in the candidate item can be expressed as: %as Eq.~(\ref{eq:transformer}):
\begin{equation}
 \label{eq:transformer}
  interest_{GM} = MHTA(\mathbf{e}^\mathbf{i},\mathbf{e}_\mathbf{g})
\end{equation}
where $\mathbf{e}^\mathbf{i}$ is the representation of the candidate item, $\mathbf{interest_{GM}}$ is the unbiased and comprehensive interest extracted from the full lifelong behavior sequence. To summarize, we believe that the computation efficiency results from reducing lots of redundant computations.

\begin{figure}
    \centering
    \includegraphics[width=\linewidth]{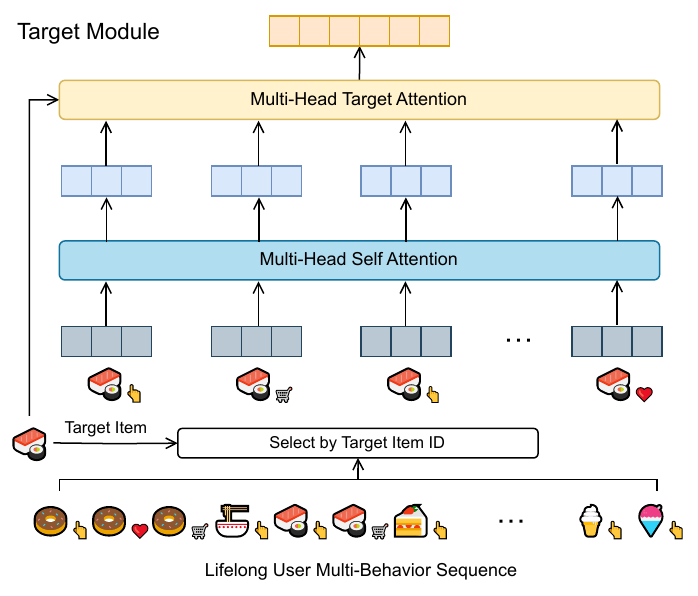}
    %\vspace{-1em}
    \caption{The detail of TM's architecture.}
    \label{fig:tm}
    %\vspace{-1em}
\end{figure}

\subsection{Target Module}
Figure~\ref{fig:tm} shows the details of TM. TM focuses on capturing the user's historical evolution process on the candidate item, which is ignored by previous methods. Firstly, TM retrieves behaviors holding the same interest key as the candidate item from the lifelong behavior sequence. The behavior pattern contained in this subsequence is a strong signal indicating the user's habit of the candidate item. For example, the user may click the candidate item every week or do a lot of micro behaviors and purchase finally. The signal is flooded to some extent in GM because other items take away some attention. However, behaviors in the subsequence are similar because of belong to the same interest key but the subsequence is short. Thus, we first use the MHSA to strengthen the subtle differences of behaviors by capturing behaviors' mutual relatedness. And then MHTA is applied to extract the psychological decision interest from the refined and differentiated behavior representation. Specifically, this candidate aware subsequence can be expressed as $\mathbf{t}=<\mathbf{t}_{1},\mathbf{t}_{2},...,\mathbf{t}_{T}>$. We use the original $\mathbf{e}^s_i$ to represent each behavior, and the subsequences representation is $\mathbf{e}_\mathbf{t}=<\mathbf{e}_{\mathbf{t}_1},\mathbf{e}_{\mathbf{t}_2},...,\mathbf{e}_{\mathbf{t}_T}> \in \mathbb{R}^{T \times {md}}$, where $m$ is the number of attributes. The modeling process follows Eq.~(\ref{eq:candidate_transformer}):
\begin{equation}
 \label{eq:candidate_transformer}
    \begin{split}
        &out^\mathbf{t}_{MHSA}=LN(\mathbf{e_t}+MHSA(\mathbf{e_t})),\\
        &out^\mathbf{t}_{enc}=LN(out^\mathbf{t}_{MHSA}+FFN(out^\mathbf{t}_{MHSA}),\\
        &out^\mathbf{t}_{MHSA}=LN(\mathbf{e}^i+MHTA(\mathbf{e}^i,out^\mathbf{t}_{enc})),\\
        &interest_{TM}=LN(out^\mathbf{t}_{MHSA}+FFN(out^\mathbf{t}_{MHSA})),
    \end{split}
\end{equation}
where $\mathbf{interest_{TM}}$ is the the psychological decision interest. 

% \subsection{Complexity Analysis}

\section{Experiment Setup}
\subsection{Datasets}
Experiments are conducted on both industrial and public datasets. The statics values of each dataset are shown in Table~\ref{tab:data_stats}.

\textbf{Industry} is the CTR dataset collected from our online LBS platform. The last 30 days' logs are used for training and samples of the following day are for testing. To exploit abundant behavior information, we collect various historical behaviors of each user from the past 2 years. The maximum length of the full lifelong behavior sequence is 10,000. There are 6 types of behaviors in our RS including click, add-to-cart, add-to-favorite, browse-dishes, view-comments, and purchase. 

\textbf{Taobao}~\cite{zhu2019joint} is a widely used dataset for CTR prediction research. It is composed of user behaviors from Taobao's industrial recommendation system. The dataset contains about 1 million users whose behaviors include clicking, adding-to-cart, and purchasing. We take the various behaviors of each user and sort them based on timestamps to construct the full lifelong behavior sequence. The maximum behavior sequence length is set to 500. Following MIMN~\cite{pi2019practice}, we use the former $T-1$ behaviors to predict whether the user will click the $T-th$ item. 
  
\begin{table}[h]
\caption{Statistics of datasets.}
\centering
\label{tab:data_stats}
%\resizebox{1.\columnwidth}{!}{
\begin{tabular}{cl|c|c|c|c}
\toprule
\multicolumn{2}{c|}{Datasets}   & \multicolumn{1}{c}{\#Users} & \multicolumn{1}{c}{\#Items} & \multicolumn{1}{c}{\#Fields} & \multicolumn{1}{c}{\#Instances} \\
\midrule
\multicolumn{2}{c|}{Taobao}           & \multicolumn{1}{c}{988K}             & \multicolumn{1}{c}{4M}            & \multicolumn{1}{c}{7}               & \multicolumn{1}{c}{0.1B}  \\
\multicolumn{2}{c|}{Industry}           & \multicolumn{1}{c}{40M}             & \multicolumn{1}{c}{417K}            & \multicolumn{1}{c}{168}               & \multicolumn{1}{c}{6.6B}  \\
\bottomrule 
\end{tabular}
%}
\end{table}
\subsection{Baselines}
We choose baselines for a comparison from four perspectives. First, we choose methods focusing on short behavior sequence modeling, including \textbf{DIN}~\cite{zhou2018deep}, \textbf{DIEN}~\cite{zhou2019deep}, \textbf{DSIN}~\cite{feng2019deep} which extracts the user's interest from short click behavior sequence. Second, we compare the proposed method with the modeling baselines over multiple types of short behavior sequences \textbf{DMT}~\cite{gu2020deep}, \textbf{DMBIN}~\cite{he2023dmbin}, \textbf{TEM4CTR}~\cite{zhang2023time}. \textbf{SIM}~\cite{pi2020search}, \textbf{ETA}~\cite{chen2022efficient}, \textbf{SDIM}~\cite{cao2022sampling},\textbf{NINN}~\cite{zhao2022non} are all methods of mining user interest from lifelong click behavior sequence, which serve as the third type of baselines. Fourth, \textbf{TWIN}~\cite{chang2023twin} concentrates on full lifelong behavior sequence modeling. We add \textbf{SIM-TM} which integrates SIM with our Target Module on category\_id to show the effectiveness of MHSA in TM. As the hierarchical attention structure of \textbf{DSIN} is also suitable for long behavior sequence modeling efficiently, we let \textbf{DSIN} directly process the long behavior sequence.

\subsection{Evaluation Metric}
Two widely used metrics AUC~\cite{bishop2006pattern}, and LogLoss are chosen. The AUC (Area Under the ROC Curve) measures the ranking accuracy. A higher AUC indicates better performance. The LogLoss measures the accuracy of the estimated probability depending on the ground-truth label. Even a slight improvement is considered a significant boost for the industry recommendation task~\cite{guo2017deepfm}, as it leads to a significant increase in revenue.

\subsection{Implementation Details}
As we observe that the user accesses the same item repeatedly on both the industrial and public datasets, we choose the item\_id as the interest key. We will show that category\_id is also a reasonable choice of interest key in the ablation study. We choose the existing concepts like item\_id and category\_id as the interest key because such choice is engineering friendly as SIM~\cite{pi2020search}. The grouping operation can be done efficiently during offline data processing. For a fair comparison, we keep the identical network structures except for the behavior sequence modeling module. For SIM, ETA, SDIM, and TWIN,  we retrieve the top 50 most candidate-relevant behaviors into the second interest extraction stage. We implement DGIN with Tensorflow. For the industry dataset, the embedding size is $16$ and the learning rate is $5e-4$. We train the model using eight $80G$ $A100$ GPUs with the batch size 1500 of a single card. For the Taobao dataset, we set the embedding size to be $18$, the learning rate to be $1e-3$, and use a single $80$ $A100$ for training with batch size $1024$. We use Adam~\cite{kingma2014adam} as the optimizer for both datasets. We run all experiments five times and report the average result. 

\section{Experiment Results}
\subsection{Overall Performance}

\begin{table}[tb!]
\centering
\caption{Performance of all methods on both datasets. The best result is in boldface and the second best is underlined. * indicates that the superiority to the best baseline is statistically significant at 0.01 level.}
\label{tab:main_result}
\begin{tabular}{cl|c|c|c|c}
\toprule
& & \multicolumn{2}{c}{Industry} &\multicolumn{2}{c}{Taobao} \\
& & \multicolumn{1}{c}{AUC} &\multicolumn{1}{c}{Logloss} &\multicolumn{1}{c}{AUC} &\multicolumn{1}{c}{Logloss} \\
\midrule
\multicolumn{2}{c|}{DIN}        & \multicolumn{1}{c}{0.6910}                & \multicolumn{1}{c|}{0.0606}                & \multicolumn{1}{c}{0.6622}                & \multicolumn{1}{c}{0.0626} \\
\multicolumn{2}{c|}{DIEN}       & \multicolumn{1}{c}{0.6916}                & \multicolumn{1}{c|}{0.0605}                & \multicolumn{1}{c}{0.6806}                & \multicolumn{1}{c}{0.0577} \\
\multicolumn{2}{c|}{DMT}        & \multicolumn{1}{c}{0.6946}                & \multicolumn{1}{c|}{0.0603}                & \multicolumn{1}{c}{0.6955}                & \multicolumn{1}{c}{0.0520} \\
\multicolumn{2}{c|}{DMBIN}      & \multicolumn{1}{c}{0.6953}                & \multicolumn{1}{c|}{0.0603}                & \multicolumn{1}{c}{0.7252}                & \multicolumn{1}{c}{0.0507} \\
\multicolumn{2}{c|}{TEM4CTR}    & \multicolumn{1}{c}{0.6959}                & \multicolumn{1}{c|}{0.0602}                & \multicolumn{1}{c}{0.7311}                & \multicolumn{1}{c}{0.0495} \\
\multicolumn{2}{c|}{SIM}        & \multicolumn{1}{c}{0.6948}                & \multicolumn{1}{c|}{0.0603}                & \multicolumn{1}{c}{0.7137}                & \multicolumn{1}{c}{0.0558} \\
\multicolumn{2}{c|}{SIM-TM}     & \multicolumn{1}{c}{0.6958}                & \multicolumn{1}{c|}{0.0602}                & \multicolumn{1}{c}{0.7256}                & \multicolumn{1}{c}{0.0500} \\
\multicolumn{2}{c|}{ETA}        & \multicolumn{1}{c}{0.6962}                & \multicolumn{1}{c|}{0.0602}                & \multicolumn{1}{c}{0.7334}                & \multicolumn{1}{c}{0.0500} \\
\multicolumn{2}{c|}{SDIM}       & \multicolumn{1}{c}{0.6979}                & \multicolumn{1}{c|}{0.0601}                & \multicolumn{1}{c}{0.7355}                & \multicolumn{1}{c}{0.0498} \\ 
\multicolumn{2}{c|}{NINN}       & \multicolumn{1}{c}{0.6965}                & \multicolumn{1}{c|}{0.0602}                & \multicolumn{1}{c}{0.7017}                & \multicolumn{1}{c}{0.0515} \\ 
\multicolumn{2}{c|}{DSIN}       & \multicolumn{1}{c}{0.6932}                & \multicolumn{1}{c|}{0.0604}                & \multicolumn{1}{c}{0.7286}                & \multicolumn{1}{c}{0.0503} \\ 
\multicolumn{2}{c|}{TWIN}       & \multicolumn{1}{c}{\underline{0.6984}}    & \multicolumn{1}{c|}{\underline{0.0600}}    & \multicolumn{1}{c}{\underline{0.7394}}    & \multicolumn{1}{c}{\underline{0.0493}}  \\ \midrule
\multicolumn{2}{c|}{DGIN}       & \multicolumn{1}{c}{\textbf{0.7028}$^*$}   & \multicolumn{1}{c|}{\textbf{0.0598}$^*$}   & \multicolumn{1}{c}{\textbf{0.7663}$^*$}   & \multicolumn{1}{c}{\textbf{0.0488}$^*$} \\ 
\bottomrule 
\end{tabular}
\end{table}

Table~\ref{tab:main_result} shows the results of all methods. DGIN obtains the best performance in both the Industry and Taobao datasets, which shows the effectiveness of DGIN. There are some insightful findings from the results. 
(1) The proposed DGIN reaches the best performance on both datasets. Compared with existing user behavior sequence modeling methods, DGIN can extract comprehensive un-biased interest and psychological decision interest from the lifelong behavior sequence in an end-to-end manner. Both two interests achieve more fine-grained user understanding, which contributes the performance improvement. 
(2) DIEN performs better than DIN, which indicates the necessity of temporal information in extracting the user's interest. 
(3) The performance is significantly improved by modeling multiple types of behavior sequences. DMT, DMBIN, and TEM4CTR take various behavior sequences to achieve better performance than methods DIN and DIEN. Compared to click behavior sequence, multiple types of behavior sequences can comprehensively reflect the user's various interests from different perspectives.
(4) SIM-TM outperforms SIM, which indicates the effectiveness of refinement and then activation paradigm on subsequence who holds the same interest key with candidate item.
(5) Long-term interest obtained from the lifelong/long behavior sequence boosts the CTR prediction accuracy further. For example, ETA, SDIM, and TWIN gain higher AUC than the former methods of focusing on the short single/multiple types of behavior sequence(s). SIM also holds an improvement compared to those methods except for DMBIN and TEM4CTR. The result reveals that the long-term interest reflects drifting and periodicity, and this long correlation is necessary for understanding the user comprehensively, which is impossible for a short behavior sequence. 
(6) The performance-increasing trend among SIM, ETA, SDIM, NINN, and TWIN demonstrates finer granularity end-to-end training is effective. The totally two-stage method SIM performs worst. SDIM uses more precise multi-round hash collision than LSH \& Hamming distance in ETA to approximate the relevance truth between candidate and behaviors in an end-to-end manner, which achieves better ranking ability. TWIN performs best among baselines with the help of consistent relevance score between the two stages brought by shared efficient target attention. 
(7) DSIN encounters severe performance degradation compared to other lifelong behavior sequence modeling methods. This result indicates that taking session as interest key leads to much information loss. We think the reason is that behaviors of the same session belong to various interests and self-attention together with Bi-LSTM are hard to preserve the behaviors' heterogeneity. Our proposed DGIN really does the end-to-end full information training on the full lifelong sequence by means of reasonable interest grouping. There is hardly any information loss in the DGIN, which is the key factor for improvement.

\subsection{Ablation Study}
\begin{table}[tb!]
\centering
\caption{Results of Integrating Each Component Successively.}
\vspace{-0.4cm}
\label{tab:ablation_component}
\begin{tabular}{cl|c|c|c|c}
\toprule
& & \multicolumn{2}{c}{Industry} &\multicolumn{2}{c}{Taobao} \\
& & \multicolumn{1}{c}{AUC} &\multicolumn{1}{c}{Logloss} &\multicolumn{1}{c}{AUC} &\multicolumn{1}{c}{Logloss} \\
\midrule
\multicolumn{2}{c|}{TWIN}                       & \multicolumn{1}{c}{0.6984}            & \multicolumn{1}{c|}{0.0600}               & \multicolumn{1}{c}{0.7394}            & \multicolumn{1}{c}{0.0493} \\ \midrule
\multicolumn{2}{c|}{DGIN-simple}                & \multicolumn{1}{c}{0.6960}            & \multicolumn{1}{c|}{0.0602}               & \multicolumn{1}{c}{0.7006}            & \multicolumn{1}{c}{0.0517} \\ \midrule
\multicolumn{2}{c|}{+Statistical Attributes}    & \multicolumn{1}{c}{0.6973}            & \multicolumn{1}{c|}{0.0602}               & \multicolumn{1}{c}{0.7364}            & \multicolumn{1}{c}{0.0500} \\
\multicolumn{2}{c|}{+Aggregated Attribute}     & \multicolumn{1}{c}{0.6998}            & \multicolumn{1}{c|}{0.0600}               & \multicolumn{1}{c}{0.7423}            & \multicolumn{1}{c}{0.0495} \\
\multicolumn{2}{c|}{+TM (DGIN)}     & \multicolumn{1}{c}{\textbf{0.7028}}   & \multicolumn{1}{c|}{\textbf{0.0598}}      & \multicolumn{1}{c}{\textbf{0.7663}}   & \multicolumn{1}{c}{\textbf{0.0488}} \\
\bottomrule 
\end{tabular}
\end{table}
In this section, we investigate the effect of each component in DGIN and display the result in Table~\ref{tab:ablation_component}. The best baseline TWIN is for comparison. All the variants stem from DGIN-simple. DGIN-simple just groups the lifelong behavior sequence and only keeps the identity attributes, which abandons lots of information. We integrate statistical attributes, aggregated attributes, and the candidate-aware subsequence to the DGIN one after another. From Table~\ref{tab:ablation_component}, we can find that all three components are beneficial for CTR prediction on both datasets.  The contributions of the interest intensity contained in statistical attributes, the temporal-space information involved in aggregated attributes, and the psychological decision interest brought by candidate-aware subsequence to performance improvement are not mutually exclusive. When we add each of them into the DGIN-simple successively, the effectiveness gains improvement step by step. The results indicate that the interests hidden in the lifelong behavior sequence are multifarious. It's necessary to design dedicated features or algorithms to capture the corresponding interest character. That is what DGIN does.  

\subsection{The Effect of Multiple Types of Behaviors}
In this section, we explore the influence of introducing multiple types of behaviors. The click-only means that we apply the DGIN to the lifelong click sequence and the \emph{behavior\_type} related attributes are removed. From Table~\ref{tab:ablation_behavior}, We have two findings. First, the click-only variant of DGIN still outperforms the strong competitor TWIN. This comparison verifies that the information exploitation of one-stage is better than the two-stage methods. Second, DGIN beats the click-only variant on both datasets. Multiple types of behaviors indeed empower the CTR model to understand the user's preference more fine-grained and comprehensively.

\begin{table}[tb!]
\centering
\caption{AUC and Logloss of modeling different types of behavior sequences on both datasets.}
\label{tab:ablation_behavior}
\begin{tabular}{cl|c|c|c|c}
\toprule
& & \multicolumn{2}{c}{Industry} &\multicolumn{2}{c}{Taobao} \\
& & \multicolumn{1}{c}{AUC} &\multicolumn{1}{c}{Logloss} &\multicolumn{1}{c}{AUC} &\multicolumn{1}{c}{Logloss} \\
\midrule
\multicolumn{2}{c|}{TWIN}               & \multicolumn{1}{c}{0.6984}            & \multicolumn{1}{c|}{0.0600}               & \multicolumn{1}{c}{0.7394}            & \multicolumn{1}{c}{0.0493} \\ \midrule
\multicolumn{2}{c|}{click-only}         & \multicolumn{1}{c}{0.7003}            & \multicolumn{1}{c|}{0.0600}               & \multicolumn{1}{c}{0.7442}            & \multicolumn{1}{c}{0.0497} \\
\multicolumn{2}{c|}{DGIN}               & \multicolumn{1}{c}{\textbf{0.7028}}   & \multicolumn{1}{c|}{\textbf{0.0598}}      & \multicolumn{1}{c}{\textbf{0.7663}}   & \multicolumn{1}{c}{\textbf{0.0488}} \\
\bottomrule 
\end{tabular}
\end{table}

\subsection{The Choice of Interest Key}
We investigate the choice of interest key to performance in this section. We first use session as interest key and then coarse-grained category\_id. We treat different item\_ids within each group when grouping on category\_id also as spatio-temporal attribute and use MHSA to aggregate them. Meanwhile, we also add some new statistical attributes including the total number of item\_id and the number of unique item\_id. To analyze the impact of the interest key on each component of DGIN thoroughly, we include the result of DGIN's variant which has no TM represented as "wo TM". The result is shown in Table~\ref{tab:ablation_groupkey} and we have the following observations. DSIN, taking session as the interest key, performs worst because the heterogeneity within each session is hard to model. Both the DGIN and its variant gain better results than TWIN, which shows the robustness of DGIN. The performance drop can be alleviated by supplementing informative human-designed attributes. On the other hand, item\_id performs better than category\_id as the grouping key. The result indicates that fine granularity grouping retains more information and benefits the latter interest extraction.

\begin{table}[tb!]
\centering
\caption{AUC and Logloss of taking different interest key.}
\label{tab:ablation_groupkey}
\begin{tabular}{cl|c|c|c|c}
\toprule
& & \multicolumn{2}{c}{Industry} &\multicolumn{2}{c}{Taobao} \\
& & \multicolumn{1}{c}{AUC} &\multicolumn{1}{c}{Logloss} &\multicolumn{1}{c}{AUC} &\multicolumn{1}{c}{Logloss} \\
\midrule
\multicolumn{2}{c|}{TWIN}               & \multicolumn{1}{c}{0.6984}            & \multicolumn{1}{c|}{0.0600}             & \multicolumn{1}{c}{0.7394}            & \multicolumn{1}{c}{0.0493} \\ \midrule
\multicolumn{2}{c|}{session (DSIN)}            & \multicolumn{1}{c}{0.6932}            & \multicolumn{1}{c|}{0.0604}             & \multicolumn{1}{c}{0.7286}            & \multicolumn{1}{c}{0.0503} \\
\multicolumn{2}{c|}{category\_id/wo TM} & \multicolumn{1}{c}{0.6985}            & \multicolumn{1}{c|}{0.0601}             & \multicolumn{1}{c}{0.7408}            & \multicolumn{1}{c}{0.0496} \\
\multicolumn{2}{c|}{category\_id}       & \multicolumn{1}{c}{0.7015}            & \multicolumn{1}{c|}{0.0599}             & \multicolumn{1}{c}{0.7639}            & \multicolumn{1}{c}{0.0490} \\
\multicolumn{2}{c|}{item\_id/wo TM}     & \multicolumn{1}{c}{0.6998}            & \multicolumn{1}{c|}{0.0600}             & \multicolumn{1}{c}{0.7423}            & \multicolumn{1}{c}{0.0495} \\
\multicolumn{2}{c|}{item\_id (DGIN)}           & \multicolumn{1}{c}{\textbf{0.7028}}   & \multicolumn{1}{c|}{\textbf{0.0598}}    & \multicolumn{1}{c}{\textbf{0.7663}}   & \multicolumn{1}{c}{\textbf{0.0488}} \\
\bottomrule 
\end{tabular}
\end{table}

\subsection{Deployment}

\begin{figure}[t]
    \centering
    \includegraphics[width=0.9\linewidth]{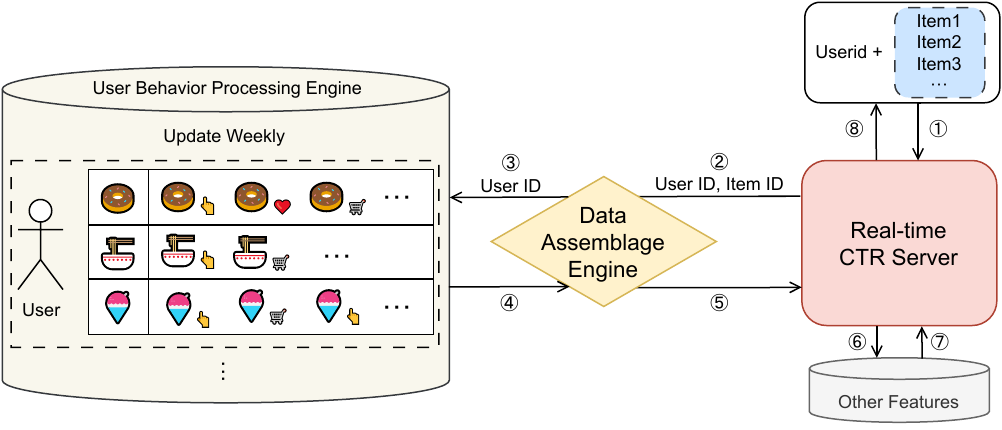}
    \vspace{-1em}
    \caption{The whole system architecture for deploying DGIN.}
    \label{fig:system}
\end{figure}

The system for deploying DGIN as shown in Figure~\ref{fig:system} contains three sub-systems: data processing, offline training, and online serving. 

\textbf{Data Processing:} We decouple the lifelong behavior sequence processing from the other features due to its requirements of storage and latency, named \textbf{U}ser \textbf{B}ehavior \textbf{P}rocessing \textbf{E}ngine (\textbf{UBPE}). UBPE collects the user's various behaviors during the past two years and then groups them based on the item\_id. We build a two-level index structure for storing the grouped full lifelong behavior sequence. The first level key is user\_id and the second level key is item\_id. UBPE will calculate the statistic values and discretize them within each group, which can free the latency of processing statistical attributes in the model stage. As the long-term interest of the user remains stable in a short period, UBPE updates the data once a week. When a new week of data comes, the UBPE distributes each behavior based on the two-level keys and then updates the statistic values in a streaming manner.

\textbf{offline training:} During the offline training stage, each training instance contains user\_id and item\_id. The data assemblage engine retrieves the corresponding grouped full lifelong behavior sequence based on the given keys. The two-level index structure allows high throughput query, which will not bottleneck the model training. 

\textbf{Online Serving:} The computation bottleneck of DGIN lies in the GM as it needs to process more behaviors. Luckily, the unbiased and comprehensive interest $\mathbf{interest_{GM}}$ can be extracted quickly by precomputing and caching the interest groups' representation $\mathbf{e}_\mathbf{g}$ after finishing the CTR model's updating each day. When a batch of candidates comes, the inference engine only needs to retrieve the candidate-aware subsequence and obtain the psychological decision interest $\mathbf{interest_{TM}}$.

\subsection{A/B Test on Performance and Cost}
% \begin{table}[h]
%     \caption{A/B Test of DGIN compared to SIM Hard.}\label{tab:online_result}
%     \vspace{-0.4cm}
%     \setlength{\tabcolsep}{8mm}
%     \begin{tabular}{c|c}
%     \toprule
%     Metrics            & Accumulated Gains   \\ \midrule %\noalign{\smallskip}\hline
%     CTR     & +4.5\% \\ 
%     RPM     & +2.0\%  \\ \bottomrule  % \noalign{\smallskip}\hline \hline\noalign{\smallskip}
%     \end{tabular}
% \end{table}
% We conduct an A/B test in the online LBS advertising system to measure the benefits of DGIN compared with the online baseline SIM Hard from 2023-05 to 2023-06. The DGIN is allocated with 10\% experiment serving traffic and the SIM Hard holds the 70\% main traffic. Table~\ref{tab:online_result} shows the relative promotion of CTR and Revenue Per Mille (RPM) during one month's testing. DGIN achieves $4.5\%$ and $2.0\%$ accumulated relative promotion on the CTR and RPM respectively during the A/B test period. This is a significant improvement in the online LBS advertising system and proves the effectiveness of DGIN. The parameter storage costs of SIM and DGIN are 2.85 GB and 3.00 GB respectively. 
% SIM spends avgerage inference latency 4.6 ms and DGIN is 5.4 ms. The resource cost bought by DGIN is negligible.
We conducted an A/B test in the online LBS advertising system to measure the benefits of DGIN compared with the online baseline SIM Hard from 2023-05 to 2023-06. The DGIN is allocated with 10\% experiment serving traffic and the SIM Hard holds the 70\% main traffic. The online result shows the relative promotion of CTR and Revenue Per Mille (RPM) during one month's testing. DGIN achieves $4.5\%$ and $2.0\%$ accumulated relative promotion on the CTR and RPM respectively during the A/B test period. This is a significant improvement in the online LBS advertising system and proves the effectiveness of DGIN. The parameter storage costs of SIM and DGIN are 2.85 GB and 3.00 GB respectively. 
SIM spends an average inference latency of 4.6 ms and DGIN is 5.4 ms. The resource cost bought by DGIN is negligible.

\section{Conclusion}
In this paper, we propose the DGIN for full lifelong user behavior sequence modeling in the CTR prediction task. DGIN, consisting of a Group Module and a Target Module, aims at extracting fine-grained comprehensive unbiased interest and psychological decision interest to achieve a deep understanding of the user's preference. To the best of our knowledge, DGIN is the first to achieve efficient end-to-end full lifelong user behavior sequence modeling. 
% The significant improvements in both offline and online evaluations demonstrate the superiority of our proposed DGIN method. 

\bibliographystyle{ACM-Reference-Format}
\bibliography{sample-base}

\appendix

\end{document}